\documentclass[aps,pra,twocolumn,showpacs,amsmath,amssymb,superscriptaddress,notitlepage]{revtex4-1}

\usepackage{graphicx}
\usepackage{dcolumn}
\usepackage{bm}
\usepackage{comment}

\begin{document}

\title{$\mathcal{PT}$-symmetric Rabi model: Perturbation theory}
\author{Tony E. Lee}
\affiliation{ITAMP, Harvard-Smithsonian Center for Astrophysics, Cambridge, Massachusetts 02138, USA}
\affiliation{Department of Physics, Indiana University Purdue University Indianapolis (IUPUI), Indianapolis, Indiana 46202, USA}
\author{Yogesh N. Joglekar}
\affiliation{Department of Physics, Indiana University Purdue University Indianapolis (IUPUI), Indianapolis, Indiana 46202, USA}

\date{\today}

\begin{abstract}
We study a non-Hermitian version of the Rabi model, where a two-level system is periodically driven with an imaginary-valued drive strength, leading to alternating gain and loss. In the Floquet picture, the model exhibits $\mathcal{PT}$ symmetry, which can be broken when the drive is sufficiently strong. We derive the boundaries of the $\mathcal{PT}$ phase diagram for the different resonances by doing perturbation theory beyond the rotating-wave approximation. For the main resonance, we show that the non-Hermitian analog of the Bloch-Siegert shift corresponds to maximal $\mathcal{PT}$ breaking. For the higher-order resonances, we capture the boundaries to lowest order. We also solve the regime of high frequency by mapping to the Wannier-Stark ladder. Our model can be experimentally realized in waveguides with spatially-modulated loss or in atoms with time-modulated spontaneous decay.
\end{abstract}

\maketitle

\section{Introduction}

In recent years, there has been substantial interest in systems that are symmetric under parity-time reversal ($\mathcal{PT}$). A non-Hermitian Hamiltonian can still have real eigenvalues if it is $\mathcal{PT}$ symmetric \cite{bender98}. In general, as a parameter is varied, a $\mathcal{PT}$-symmetric Hamiltonian undergoes a transition from a $\mathcal{PT}$-symmetric phase (real eigenvalues) to a $\mathcal{PT}$-broken phase (complex eigenvalues). In the $\mathcal{PT}$-symmetric phase, the system exhibits periodic oscillations in time, while in the $\mathcal{PT}$-broken phase, it exhibits exponential growth. This discovery opened up the possibility of studying what new physics arises in \mbox{non-Hermitian} systems \cite{guo09,graefe10,ruter10,uzdin11,heiss12,thompson12,lee14b,lee14d,lee15a,li15,wu15,znojil15,zeuner15,kepesidis15,bender15}.

While research on $\mathcal{PT}$ symmetry has focused on time-independent Hamiltonians, several recent works have studied time-dependent Hamiltonians \cite{luo13,joglekar14,gong15}. A recent work considered a non-Hermitian Rabi model, where a two-level system is periodically driven in time with an imaginary-valued drive strength, i.e., there is alternating gain and loss \cite{joglekar14}. There is a rich $\mathcal{PT}$ phase diagram as a function of drive frequency and strength. Furthermore, an imaginary drive can induce Rabi oscillations between the two states \cite{gong15}.

In this paper, we use perturbation theory to derive the boundaries of the $\mathcal{PT}$ phase diagram for the non-Hermitian Rabi model. In particular, we capture higher-order effects beyond the rotating-wave approximation. The model is defined as
\begin{eqnarray}
H&=&\frac{\omega_o}{2} \sigma_z + 2i\lambda \sigma_x \cos{\omega t}, \label{eq:H_nonherm}
\end{eqnarray}
which represents a two-level system with energy difference $\omega_o$, driven periodically at frequency $\omega$. The drive strength $i\lambda$ is imaginary, leading to periodic gain and loss in time. $\sigma_x$ and $\sigma_z$ are Pauli operators. The goal is to calculate, as a function of parameters, whether evolution under Eq.~\eqref{eq:H_nonherm} leads to periodic dynamics ($\mathcal{PT}$-symmetric phase) or to exponential growth ($\mathcal{PT}$-broken phase). In the Floquet picture \cite{shirley65,dittrich98}, we want to see whether the Floquet quasienergies of Eq.~\eqref{eq:H_nonherm} are real or complex. The strategy here is to recall the known results for the Hermitian Rabi model (with a real-valued drive strength), and then let the drive strength become imaginary and see when the Floquet quasienergies become complex.

\begin{figure}[t]
\centering
\includegraphics[width=3.5 in,trim=1.3in 3.1in 1.6in 3.5in,clip]{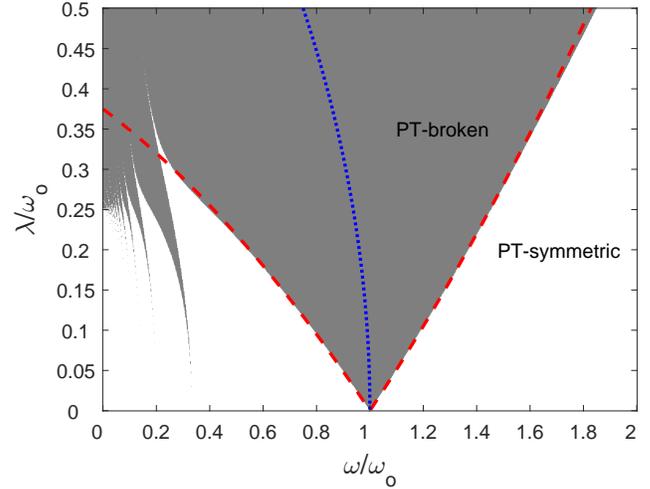}
\caption{\label{fig:zoommid} (Color online) $\mathcal{PT}$ phase diagram, showing $\mathcal{PT}$-symmetric (white) and $\mathcal{PT}$-broken (gray) phases. Also shown are the boundaries for the single-photon resonance (red dashed line) and Bloch-Siegert shift (blue dotted line) calculated using perturbation theory at next-to-leading order.}
\end{figure}

\begin{figure}[t]
\centering
\includegraphics[width=3.5 in,trim=1.3in 3.1in 1.6in 3.5in,clip]{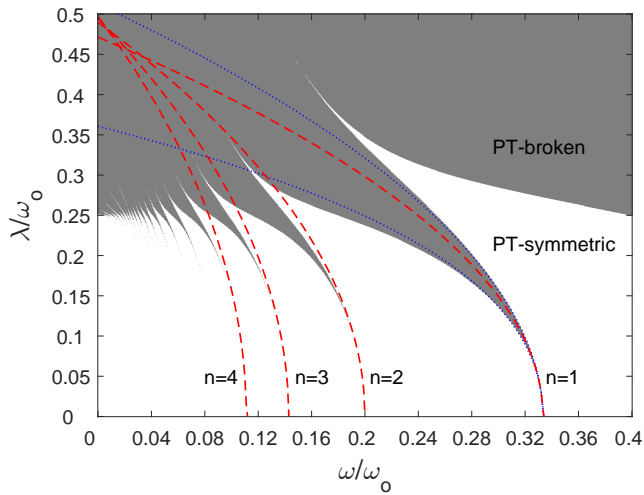}
\caption{\label{fig:zoomin} (Color online) Zoomed-in view of $\mathcal{PT}$ phase diagram, showing multi-photon resonances at $\omega\approx\omega_o/(2n+1)$. Also shown are the lowest-order boundaries for some multi-photon resonances (red dashed lines) calculated using perturbation theory. The next-order boundaries for $n=1$ are also shown (blue dotted lines).}
\end{figure}

We first derive the boundary for the ``single-photon'' resonance (Fig.~\ref{fig:zoommid}). In particular, we find the curvature due to counter-rotating terms and identify the non-Hermitian analog of the Bloch-Siegert shift \cite{bloch40}. Then we derive the boundaries of the ``multi-photon'' resonances to lowest order (Fig.~\ref{fig:zoomin}) and show that the width of successive resonances decreases exponentially. We also derive the boundaries in the limits of low and high frequencies, where for the latter, we use a modified high-frequency Floquet analysis.

Equation \eqref{eq:H_nonherm} can be implemented in experiments with waveguides, where propagation in space corresponds to propagation in time \cite{garanovich12}. Recent waveguide experiments have implemented the Hermitian Rabi model \cite{shandarova09}. By adding a gain-loss profile that is modulated in space, one obtains Eq.~\eqref{eq:H_nonherm}. Actually, it is not necessary to include gain: if there is only modulated loss, one obtains Eq.~\eqref{eq:H_nonherm} on top of a background of decay, and the $\mathcal{PT}$-symmetric phase corresponds to slower decay \cite{guo09}. Alternatively, one can implement the model with a trapped atom by modulating the spontaneous emission rate (by modulating the optical pumping) and post-selecting on trials without decay events \cite{lee14b,lee14d,lee15a}.

In Sec.~\ref{sec:herm}, we review the perturbation theory for the Hermitian Rabi model. In Sec.~\ref{sec:nonherm}, we examine the non-Hermitian model, considering the single-photon resonance, then the multi-photon resonances, and finally the limits of low and high frequency. The Appendixes provide details of the perturbation theory.

\section{Review of Hermitian Rabi model} \label{sec:herm}

First, we briefly review the Hermitian Rabi model,
\begin{eqnarray}
H&=&\frac{\omega_o}{2} \sigma_z + 2\lambda \sigma_x \cos{\omega t},
\end{eqnarray}
where the drive strength $\lambda$ is real. We work in the $\sigma_z$ basis: $\left|\uparrow\right\rangle, \left|\downarrow\right\rangle$. This model was solved perturbatively in $\lambda$ in Ref.~\cite{shirley65} using the perturbation theory of Salwen \cite{salwen55}.


We move to the Floquet picture \cite{shirley65,dittrich98}. We define the basis $|\alpha n\rangle$ for the Floquet Hamiltonian, where $\alpha=\uparrow,\downarrow$ and $n$ is any integer. The Floquet Hamiltonian $H_F$ is an infinite-dimensional matrix given by \cite{shirley65,dittrich98}
\begin{widetext}
\begin{eqnarray}
\langle \alpha n|H_F|\beta m\rangle &=& \langle \alpha |\left(-\frac{\omega_o}{2}\sigma_z + n\omega I\right)\delta_{nm} + \lambda\sigma_x(\delta_{m,n+1} + \delta_{m,n-1})|\beta \rangle.
\end{eqnarray}
For example, the block of $H_F$ corresponding to $\left|\uparrow -1\right\rangle$, $\left|\downarrow -1\right\rangle$, $\left|\uparrow 0\right\rangle$, $\left|\downarrow 0\right\rangle$, $\left|\uparrow 1\right\rangle$, $\left|\downarrow 1\right\rangle$, $\left|\uparrow 2\right\rangle$, $\left|\downarrow 2\right\rangle$ is
\begin{eqnarray}
H_F&=&\left(
\begin{array}{cccccccccc}
\cdot &         \cdot             &          \cdot             &       \cdot        &      \cdot          &          \cdot            &        \cdot               &        \cdot               &         \cdot              & \cdot  \\
\cdot & \frac{\omega_o}{2}-\omega &            0               &         0          &  \lambda            &           0               &          0                 &           0                &            0               & \cdot  \\  
\cdot &          0                & -\frac{\omega_o}{2}-\omega &   \lambda          &        0            &           0               &          0                 &           0                &            0               & \cdot  \\
\cdot &          0                &        \lambda             & \frac{\omega_o}{2} &        0            &           0               &        \lambda             &           0                &            0               & \cdot  \\ 
\cdot &  \lambda                  &            0               &        0           & -\frac{\omega_o}{2} &      \lambda              &          0                 &           0                &            0               & \cdot  \\
\cdot &          0                &            0               &        0           &    \lambda          & \frac{\omega_o}{2}+\omega &          0                 &           0                &         \lambda            & \cdot  \\
\cdot &          0                &            0               &    \lambda         &        0            &           0               & -\frac{\omega_o}{2}+\omega &         \lambda            &            0               & \cdot  \\
\cdot &          0                &            0               &        0           &        0            &           0               &         \lambda            & \frac{\omega_o}{2}+2\omega &            0               & \cdot  \\
\cdot &          0                &            0               &        0           &        0            &      \lambda              &          0                 &           0                & -\frac{\omega_o}{2}+2\omega& \cdot  \\
\cdot &         \cdot             &          \cdot             &       \cdot        &      \cdot          &         \cdot             &         \cdot              &         \cdot              &           \cdot            & \cdot  
\end{array}
\right). \label{eq:HF_matrix}
\end{eqnarray}
\end{widetext}

We seek the eigenvalues $\epsilon$ of $H_F$, since they are the quasienergies of $H$. We are mainly interested in when $\omega$ is such that two diagonal elements of $H_F$ are nearly degenerate, since this corresponds to a resonance. The main (single-photon) resonance occurs when $\omega\approx\omega_o$. The multi-photon resonances occur when $\omega_o\approx 3\omega,5\omega, 7\omega,\ldots$, i.e., when $\omega_o$ is an odd multiple of $\omega$.

When two diagonal elements of $H_F$ are nearly degenerate, we use Salwen's perturbation theory \cite{salwen55} to calculate the effective $2\times 2$ Hamiltonian for this subspace. The effective Hamiltonian captures the effect of all other states on the two degenerate states. Once we have this effective Hamiltonian, we diagonalize it to obtain $\epsilon$. In Appendix \ref{sec:salwen}, we briefly review Salwen's perturbation theory, since it may be unfamiliar to many readers.

\subsection{Single-photon resonance: $\omega\approx\omega_o$} \label{sec:herm_single}

Suppose that $\omega\approx\omega_o$, such that the unperturbed states $\left|\uparrow 0\right\rangle$ and $\left|\downarrow 1\right\rangle$ are almost degenerate. Using Salwen's perturbation theory (see Appendix \ref{sec:salwen}), we find the effective $2\times 2$ Hamiltonian $H_F'$ for these two states as a perturbation series of $\lambda$. One must be careful when doing the perturbation theory: $(\omega-\omega_o)$ and $(\epsilon-\omega_o/2)$ are assumed to be on the order of $\lambda$.

To lowest order, the effective Hamiltonian is \cite{shirley65}
\begin{eqnarray}
H_F'&=&\left(\begin{array}{cc}
\frac{\omega_o}{2}   &   \lambda                         \\
\lambda              &  -\frac{\omega_o}{2} + \omega
\end{array}\right),
\end{eqnarray}
which has eigenvalues,
\begin{eqnarray}
\epsilon&=& \frac{\omega \pm \Omega}{2}, \label{eq:herm_epsilon_single_lowest_order}\\
\Omega^2&=& (\omega-\omega_o)^2 + 4\lambda^2,
\end{eqnarray}
where $\Omega$ is the generalized Rabi frequency, i.e., the frequency of the Rabi oscillations. This lowest-order approximation is the ``rotating-wave approximation,'' since it ignores the coupling between $\left|\uparrow 0\right\rangle$ and $\left|\downarrow -1\right\rangle$ and between $\left|\downarrow 1\right\rangle$ and $\left|\uparrow 2\right\rangle$, which are the ``counter-rotating'' terms.

To next order, the effective Hamiltonian is \cite{shirley65}
\begin{eqnarray}
H_F'&=&\left(\begin{array}{cc}
\frac{\omega_o}{2} + \frac{\lambda^2}{2\omega_o}  &   \lambda                         \\
\lambda                                           &  -\frac{\omega_o}{2} + \omega - \frac{\lambda^2}{2\omega_o}
\end{array}\right),
\end{eqnarray}
where there is now a level shift due to coupling to $\left|\downarrow -1\right\rangle$ and $\left|\uparrow 2\right\rangle$. The eigenvalues are now
\begin{eqnarray}
\epsilon&=& \frac{\omega \pm \tilde{\Omega}}{2}, \label{eq:herm_epsilon_single_next_order}\\
\tilde{\Omega}^2&=&(\omega-\omega_o)^2 + 4\lambda^2 -\frac{2(\omega-\omega_o)\lambda^2}{\omega_o},	\label{eq:Omega_single}
\end{eqnarray}
where $\tilde{\Omega}$ is the effective Rabi frequency. We have truncated $\tilde{\Omega}^2$ at $O(\lambda^3)$, because it is only accurate to that order. [Improving the accuracy would require calculating the level shift beyond $O(\lambda^2)$, which has been done \cite{dittrich98,aravind84}, but we stop here for simplicity.]

From Eq.~\eqref{eq:Omega_single}, one finds a shift in the resonance frequency $\omega_\text{res}$, defined as the frequency at which the Rabi oscillations have maximum amplitude. Since the amplitude of Rabi oscillations is proportional to $1/\tilde{\Omega}^2$ \cite{dittrich98},  the oscillations achieve maximum amplitude when $d\tilde{\Omega}^2/d\omega=0$:
\begin{eqnarray}
\omega_\text{res}&=&\omega_o + \frac{\lambda^2}{\omega_o}.
\end{eqnarray}
This is the famous Bloch-Siegert shift, which is due to the counter-rotating terms \cite{shirley65, dittrich98, bloch40}.

\subsection{Multi-photon resonance: $\omega_o\approx(2n+1)\omega$}  \label{sec:herm_multi}

Now suppose that $\omega_o\approx(2n+1)\omega$ where $n$ is a positive integer. Then the unperturbed states, $\left|\uparrow 0\right\rangle$ and $\left|\downarrow (2n+1)\right\rangle$, are almost degenerate. From Salwen's perturbation theory, we obtain the effective $2\times 2$ Hamiltonian for these two states:
\begin{eqnarray}
H_F'&=&\left(\begin{array}{cc}
\frac{\omega_o}{2} + \delta   &   u                         \\
u                             &  -\frac{\omega_o}{2} + (2n+1)\omega - \delta
\end{array}\right),
\end{eqnarray}
where $\delta$ is the level shift, and $u$ is the effective coupling between the states. Note that there is no direct coupling between $\left|\uparrow 0\right\rangle$ and $\left|\downarrow (2n+1)\right\rangle$ in Eq.~\eqref{eq:HF_matrix}, so $u$ arises from coupling to intermediate states. The eigenvalues are
\begin{eqnarray}
\epsilon&=& \frac{(2n+1)\omega \pm \tilde{\Omega}}{2}, \label{eq:herm_epsilon_multi}\\
\tilde{\Omega}^2&=& [(2n+1)\omega-\omega_o-2\delta]^2 + 4u^2, \label{eq:Omega_multi}
\end{eqnarray}

One finds \cite{shirley65, aravind84}
\begin{eqnarray}
\delta&=&\frac{2n+1}{2n(n+1)}\frac{\lambda^2}{\omega}, \label{eq:delta_multi}\\
u&=&\frac{(-1)^n \lambda^{2n+1}}{2^{2n}(n!)^2\omega^{2n}}, \label{eq:u_multi}
\end{eqnarray}
where Eq.~\eqref{eq:delta_multi} is accurate to $O(\lambda^2)$, and Eq.~\eqref{eq:u_multi} is accurate to $O(\lambda^{2n+1})$. Equation \eqref{eq:u_multi} shows that the larger $n$ is, the weaker the transition, as expected.



\section{Non-Hermitian Rabi Model} \label{sec:nonherm}

\begin{figure}[b]
\centering
\includegraphics[width=3.5 in,trim=1.3in 3.1in 1.6in 3.5in,clip]{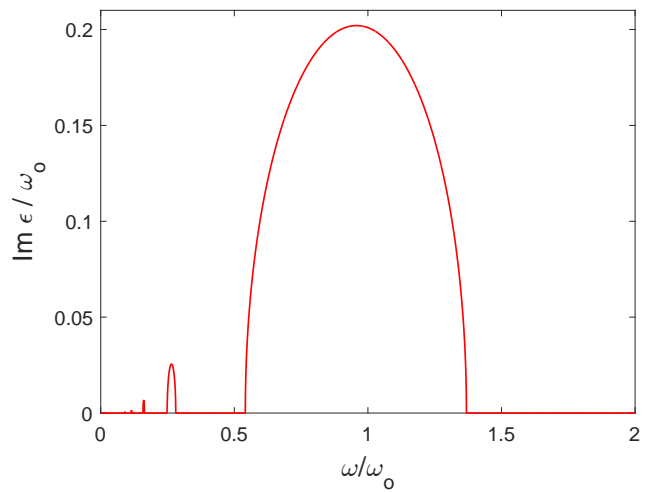}
\caption{\label{fig:eigenvalues} (Color online) Imaginary part of Floquet quasienergy $\epsilon$ for $\lambda=0.2\omega_o$, showing windows of $\mathcal{PT}$ breaking.}
\end{figure}

Now we adapt the results of Sec.~\ref{sec:herm} to the non-Hermitian case by replacing $\lambda\rightarrow i\lambda$ throughout. Although $H_F$ in Eq.~\eqref{eq:HF_matrix} becomes non-Hermitian, it is $\mathcal{PT}$-symmetric using $\mathcal{P}=\sigma_z \otimes I$, so we expect the eigenvalues $\epsilon$ to be real for small $\lambda$ and complex for large $\lambda$. (Although $H_F$ has an infinite number of eigenvalues, they all become complex simultaneously, so we only need to track one of them.)

Figures \ref{fig:zoommid} and \ref{fig:zoomin} show that there are windows of $\mathcal{PT}$ breaking whenever $\omega$ hits a resonance. The intuitive reason for this is that when the drive frequency is close to a resonance, the imaginary drive has more effect and thus breaks $\mathcal{PT}$ symmetry. On the other hand, when the drive is not close to a resonance, it does not have much effect and $\mathcal{PT}$ symmetry is maintained. Figure \ref{fig:eigenvalues} plots the imaginary part of $\epsilon$ as a function of $\omega$.

We seek to derive the boundary between the $\mathcal{PT}$-symmetric and $\mathcal{PT}$-broken phases. It is clear from Eqs.~\eqref{eq:herm_epsilon_single_lowest_order}, \eqref{eq:herm_epsilon_single_next_order}, and \eqref{eq:herm_epsilon_multi} that $\epsilon$ becomes complex when $\Omega^2$ or $\tilde{\Omega}^2$ becomes negative. Thus, the boundary between the $\mathcal{PT}$-symmetric and $\mathcal{PT}$-broken phases is given by when $\Omega^2$ or $\tilde{\Omega}^2$ is zero.

\subsection{Single-photon resonance: $\omega\approx\omega_o$}  \label{sec:nonherm_single}

We first consider the case of $\omega\approx\omega_o$, replacing $\lambda\rightarrow i\lambda$ in Sec.~\ref{sec:herm_single}. We still call this the ``single-photon'' resonance even though the imaginary drive does not involve photons. To lowest order, the eigenvalues are
\begin{eqnarray}
\epsilon&=& \frac{\omega \pm \Omega}{2}, \label{eq:nonherm_epsilon_single_lowest_order}\\
\Omega^2&=& (\omega-\omega_o)^2 - 4\lambda^2.
\end{eqnarray}
The system is in the $\mathcal{PT}$-broken phase ($\Omega^2<0$) when
\begin{eqnarray}
\lambda &>& \frac{|\omega-\omega_o|}{2}.
\end{eqnarray}
Thus, the boundary is linear under the rotating-wave approximation. This was derived previously in Ref.~\cite{joglekar14}.

\begin{figure*}[t]
\centering
\includegraphics[width=7 in,trim=0.2in 4.1in 0.2in 4.1in,clip]{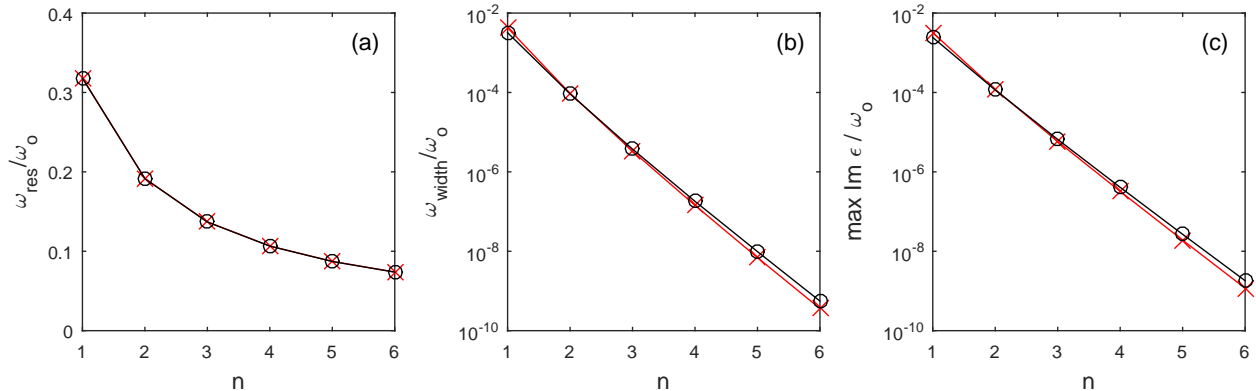}
\caption{\label{fig:multi_properties} (Color online)  Properties of multi-photon resonances at $\lambda=0.1\omega_o$ for different $n$, calculated numerically (black circles) and via rough perturbation theory (red crosses). (a) Frequency where maximal $\mathcal{PT}$ breaking occurs. (b) Width of $\mathcal{PT}$ breaking window. (c) Maximum imaginary part of Floquet quasienergy, which determines the maximum growth rate.}
\end{figure*}

To next order, 
\begin{eqnarray}
\epsilon&=& \frac{\omega \pm \tilde{\Omega}}{2}, \label{eq:nonherm_epsilon_single_next_order}\\
\tilde{\Omega}^2&=&(\omega-\omega_o)^2 - 4\lambda^2 + \frac{2(\omega-\omega_o)\lambda^2}{\omega_o}.
\end{eqnarray}
So the system is in the $\mathcal{PT}$-broken phase when
\begin{eqnarray}
\lambda &>& \frac{|\omega-\omega_o|}{2}\left(1 + \frac{\omega-\omega_o}{4\omega_o}\right), \label{eq:nonherm_single_next_boundary}
\end{eqnarray}
where we have truncated at $O(\lambda^2)$, which is the accuracy at this level. Thus, we have obtained the quadratic correction to the boundary due to the counter-rotating terms. This agrees well with numerics for small $\lambda$ (Fig.~\ref{fig:zoommid}).

In the non-Hermitian case, there is still a Bloch-Siegert shift. When $\epsilon$ is complex, the evolution under $H$ leads to exponential growth, and the growth rate is maximum when $\tilde{\Omega}^2$ reaches its minimum, which happens at
\begin{eqnarray}
\omega_\text{res}&=&\omega_o - \frac{\lambda^2}{\omega_o}.
\end{eqnarray}
At this point, the imaginary part of $\epsilon$ reaches its maximum. Thus, the Bloch-Siegert shift corresponds to the point at which the solutions exhibit the fastest exponential growth. Note that due to the $i^2$, the Bloch-Siegert shift for the non-Hermitian case is in the opposite direction of the Hermitian case.

\subsection{Multi-photon resonance: $\omega_o\approx(2n+1)\omega$}  \label{sec:nonherm_multi}

At the multi-photon resonances, $\omega_o\approx(2n+1)\omega$, there are windows of $\mathcal{PT}$ breaking in the phase diagram (Fig.~\ref{fig:zoomin}). Here we obtain the windows to lowest order. Adapting the results of Sec.~\ref{sec:herm_multi} to the non-Hermitian case:
\begin{eqnarray}
\epsilon&=& \frac{(2n+1)\omega \pm \tilde{\Omega}}{2}, \label{eq:nonherm_epsilon_multi}\\
\tilde{\Omega}^2&=& [(2n+1)\omega-\omega_o-2\delta]^2 + 4u^2, \label{eq:nonherm_Omega_multi} \\
\delta&=&-\frac{2n+1}{2n(n+1)}\frac{\lambda^2}{\omega}, \label{eq:nonherm_delta_multi}\\
u&=&\frac{i \lambda^{2n+1}}{2^{2n}(n!)^2\omega^{2n}}. \label{eq:nonherm_u_multi}
\end{eqnarray}
We see that at lowest order, $\tilde{\Omega}^2$ does not depend on $u$. However, when we omit $u$ from Eq.~\eqref{eq:nonherm_Omega_multi}, $\tilde{\Omega}^2$ cannot become negative; its minimum possible value is zero. This means that at lowest order, we can only get a line in parameter space corresponding to when $\mathcal{PT}$ symmetry is on the verge of breaking.

The line is given by when $(2n+1)\omega-\omega_o-2\delta=0$, or
\begin{eqnarray}
\lambda(n)&=& \left[-\frac{n(n+1)\omega_o}{2n+1}\left(\omega-\frac{\omega_o}{2n+1}\right)\right]^{1/2}, \label{eq:line_multi}
\end{eqnarray}
where we have used the fact that $\omega_o\approx(2n+1)\omega$.  For any $n$, the line has a square-root shape. As $n$ increases, the line becomes steeper.

This result agrees well with numerics (Fig.~\ref{fig:zoomin}). For small $\lambda$, each window is very narrow and follows the predicted line. In fact, Fig.~\ref{fig:multi_properties}(a) shows that Eq.~\eqref{eq:line_multi} is useful for predicting the frequency where maximal $\mathcal{PT}$ breaking occurs (where $\text{Im }\epsilon$ is maximum). In analogy with the single-photon case, we label this $\omega_\text{res}$.

In order to go further and predict an area for each window, we would have to go to high-enough order in perturbation theory such that $u$ is included in $\tilde{\Omega}^2$. This means we would have to obtain $\delta$ to $O(\lambda^{4n})$, which can be tedious. In Appendix \ref{sec:app_three}, we do this for $n=1$.

As a rough approximation for the windows, we can simply use the leading-order value of $\delta$ given in Eq.~\eqref{eq:nonherm_delta_multi}. Although this is not a rigorous perturbative expansion, it does give reasonable estimates for the window boundaries, which are found by solving $(2n+1)\omega-\omega_o-2\delta=\pm 2iu$. We use this approach to estimate the width of the window for each $n$. Approximating $n!\approx \sqrt{2\pi n}(n/e)^n$, the window width is
\begin{eqnarray}
\omega_\text{width}(n)&\approx&\frac{2(e\lambda)^{2n+1}}{\pi n(2n+1)\omega_o^{2n}}.
\end{eqnarray}
This agrees reasonably well with numerics [Fig.~\ref{fig:multi_properties}(b)]. The deviations are mainly due to omitting the higher-order terms in $\delta$. Thus, as $n$ increases, the width decreases exponentially. (We assume $\lambda<\omega_o/e$.)

\begin{figure}[t]
\centering
\includegraphics[width=3.5 in,clip]{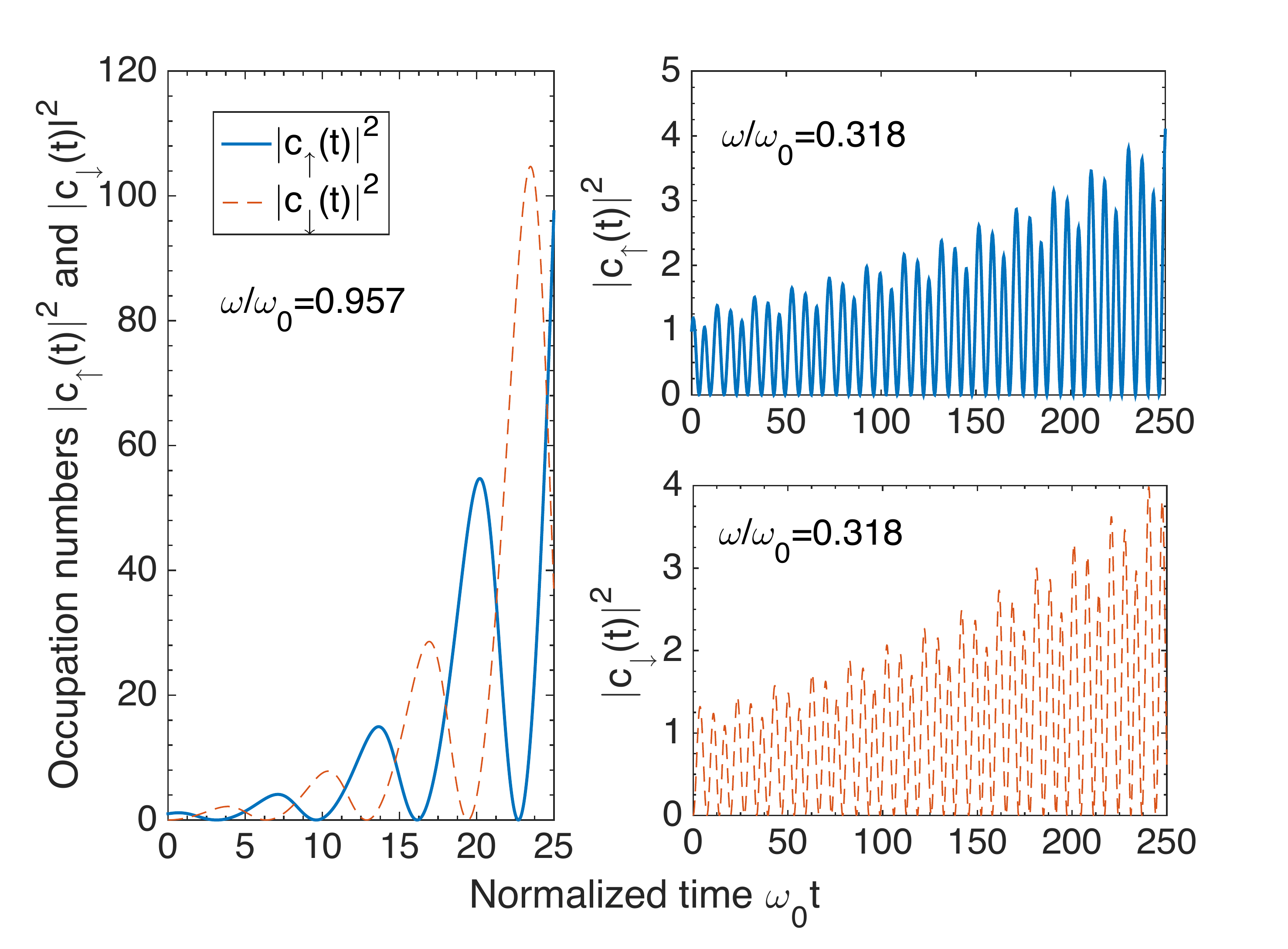}
\caption{\label{fig:trajectories} (Color online) Example solutions in the $\mathcal{PT}$-broken phase for the single-photon resonance (left) and the three-photon resonance (right) with $\lambda=0.1\omega_o$. The occupation in each state is plotted: $|c_\uparrow|^2=|\left\langle\uparrow\right|\psi\rangle|^2$ and $|c_\downarrow|^2=|\left\langle\downarrow\right|\psi\rangle|^2$.}
\end{figure}

We can similarly obtain a rough estimate for the maximum imaginary part of $\epsilon$ for each multi-photon resonance. This quantity is related to the maximum growth rate in the $\mathcal{PT}$-broken phase. We find
\begin{eqnarray}
\text{max Im }\epsilon(n)&\approx& \frac{(e\lambda)^{2n+1}}{2\pi n\omega_o^{2n}},
\end{eqnarray}
which again agrees reasonably well with numerics [Fig.~\ref{fig:multi_properties}(c)]. So as $n$ increases, the growth rate decreases exponentially. This means that in order to observe $\mathcal{PT}$ breaking for higher $n$, one must evolve the system for longer time (or longer distance in the case of waveguides). Figure \ref{fig:trajectories} shows example trajectories in the $\mathcal{PT}$-broken phase for the single-photon resonance and the three-photon resonance ($n=1$): the latter clearly grows slower than the former.

\subsection{Low frequency: $\omega\ll \omega_o$}  \label{sec:nonherm_lowfreq}

We now consider the limit of low frequency. It is clear from Fig.~\ref{fig:zoomin} that as $\omega\rightarrow 0$, the $\mathcal{PT}$ boundary approaches $\lambda=\omega_o/4$. This can be understood as follows.  For small $\omega$, the non-Hermitian term in Eq.~\eqref{eq:H_nonherm} changes very slowly, so the Hamiltonian is basically static. We can replace $2i\lambda\cos\omega t$ with its maximum value, $2i\lambda$, since $\mathcal{PT}$ symmetry is most likely to be broken at that point of the cycle. Thus, we have
\begin{eqnarray}
H&=&\frac{\omega_o}{2} \sigma_z + 2i\lambda \sigma_x. 
\end{eqnarray}
It is easy to show that this Hamiltonian is in the $\mathcal{PT}$-broken phase when $\lambda>\omega_o/4$, in agreement with numerics in the limit $\omega\rightarrow0$ (Fig.~\ref{fig:zoomin}). However, note that if one sets $\lambda=\omega_o/4$ and takes the limit $\omega\rightarrow 0$, one goes through an infinite series of multi-photon transitions \cite{vemuri15}.

\subsection{High frequency: $\omega\gg \omega_o$}  \label{sec:nonherm_highfreq}

\begin{figure}[t]
\centering
\includegraphics[width=3.5 in,trim=1.3in 3.1in 1.6in 3.5in,clip]{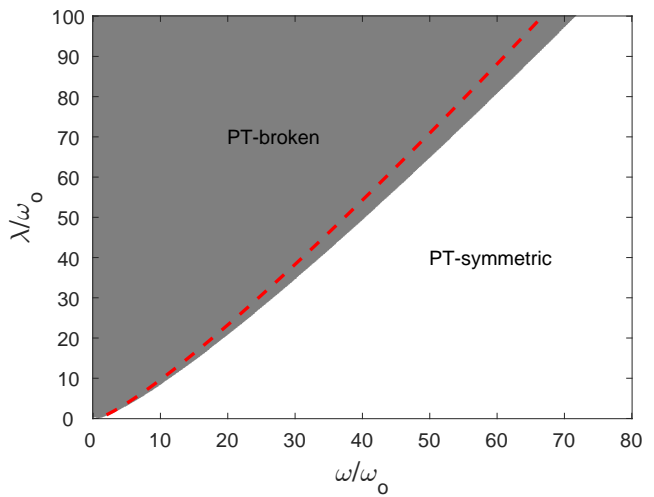}
\caption{\label{fig:zoomout} (Color online) Zoomed-out view of $\mathcal{PT}$ phase diagram. Also shown is the boundary (red dashed line) calculated using perturbation theory in $\omega_o$.}
\end{figure}

In the limit of large frequency, the $\mathcal{PT}$ boundary continues to rise (Fig.~\ref{fig:zoomout}). It is impractical to keep doing perturbation theory in $\lambda$ as in Sec.~\ref{sec:nonherm_single}. Instead, we can treat $\omega_o$ as the perturbation, and this allows us to obtain the boundary in this limit. Here, we just use conventional perturbation theory -- not Salwen's perturbation theory.

[We note that for the Hermitian Rabi model, one typically treats this limit by transforming to a rotating frame and then averaging over a period of $\omega$ to obtain an effectively unmodulated system \cite{dittrich98}. When applied to the non-Hermitian model in Eq.~\eqref{eq:H_nonherm}, this approach predicts that $\mathcal{PT}$ symmetry is never broken, which is \mbox{wrong} (but see \footnote{The averaging method does work when the modulation is in the Hermitian term instead of the non-Hermitian term \cite{luo13}.}). It turns out that averaging over a period does not work for Eq.~\eqref{eq:H_nonherm} because it discards information about large amplitude oscillations. Thus, we must use the alternative approach described below.]

It is clear from Eq.~\eqref{eq:HF_matrix} that there are two uncoupled groups of states. The first group includes $\left|\uparrow 0\right\rangle$, $\left|\downarrow 1\right\rangle$,  $\left|\uparrow 2\right\rangle$, $\left|\downarrow 3\right\rangle$. The second group includes $\left|\downarrow 0\right\rangle$,  $\left|\uparrow 1\right\rangle$, $\left|\downarrow 2\right\rangle$, $\left|\uparrow 3\right\rangle$. For the following discussion, we only consider the first group. (The second group gives the same results.)

For the first group, we can rewrite Eq.~\eqref{eq:HF_matrix} as
\begin{eqnarray}
H_F&=&H_0 + H_1,\\
H_0&=&\sum_m [i\lambda(|m\rangle\langle m+1| + |m+1\rangle\langle m|) + m\omega |m\rangle\langle m|], \nonumber\\ \label{eq:tightbinding}\\
H_1&=&\frac{\omega_o}{2} \sum_m (-1)^m |m\rangle\langle m|,
\end{eqnarray}
where we have replaced $\lambda\rightarrow i\lambda$. Let the eigenvalues and eigenstates of $H_F$ be denoted $\epsilon_n$ and $|\phi_n\rangle$, where $n$ is any integer. (Here, $n$ is unrelated to the $n$ of previous sections!) We treat $H_1$ as the perturbation.

$H_0$ is a well-known Hamiltonian: it is a tight-binding model with a linear potential and leads to the Wannier-Stark ladder \cite{gluck02,bender15}. The eigenvalues of $H_0$ are
\begin{eqnarray}
\epsilon^{(0)}_n &=& n\omega,
\end{eqnarray}
and the eigenstates can be written in terms of Bessel functions of the first kind,
\begin{eqnarray}
|\phi_n^{(0)}\rangle&=&\sum_m J_{n-m} \left(\frac{2i\lambda}{\omega}\right)|m\rangle.
\end{eqnarray}
Note that $\epsilon^{(0)}_n$ is always real.

Now we calculate the first-order shift due to $H_1$. For simplicity, we only calculate the shift for the $n=0$ eigenvalue:
\begin{eqnarray}
\epsilon^{(1)}_0 &=& \langle \phi_0^{(0)}|H_1|\phi_0^{(0)}\rangle,\\
&=&\frac{\omega_o}{2}\sum_m (-1)^m  J_{-m} \left(\frac{2i\lambda}{\omega}\right) J_{-m} \left(\frac{2i\lambda}{\omega}\right),\\
&=&\frac{\omega_o}{2}\sum_m  J_{m} \left(\frac{2i\lambda}{\omega}\right) J_{-m} \left(\frac{2i\lambda}{\omega}\right),\\
&=&\frac{\omega_o}{2} J_{0} \left(\frac{4i\lambda}{\omega}\right),
\end{eqnarray}
which is always real and positive. We have used Bessel-function identities in the last two lines. One can show that $\epsilon^{(1)}_0=-\epsilon^{(1)}_1$.

However, there is a problem. It is clear that doing perturbation theory in this way will never be able to exhibit a $\mathcal{PT}$ breaking transition, since such a transition requires two eigenvalues to become degenerate and would thus require going to infinite order. (Remember that we are not using Salwen's method.) Nonetheless, we can sidestep this issue using the following approach.

We note that the $\mathcal{PT}$ transition occurs because as $\omega_o$ increases, $\epsilon_0$ increases from zero and $\epsilon_1$ decreases from $\omega$ until the two coalesce and become complex. (Due to periodicity of Floquet eigenvalues, we only have to consider $\epsilon_0$ and $\epsilon_1$.) We can approximate when this degeneracy occurs using the fact that $\epsilon^{(0)}_1 - \epsilon^{(0)}_0=\omega$ and that $\epsilon^{(1)}_0=-\epsilon^{(1)}_1$. Thus, we expect the degeneracy to occur when $2\epsilon^{(1)}_0 = \omega$, or
\begin{eqnarray}
\frac{\omega}{\omega_o}&=& J_{0} \left(\frac{4i\lambda}{\omega}\right).
\end{eqnarray}
One has to solve this transcendental equation numerically. The solution is plotted in Fig.~\ref{fig:zoomout}, which shows reasonable agreement with numerics. The deviations are probably due to the rough way we predicted the degeneracy.

\section{Conclusion}

We have derived the $\mathcal{PT}$ phase diagram of the \mbox{non-Hermitian} Rabi model in different limits. A future direction is to extend the results to a many-body setting, where the imaginary drive couples to many interacting spins instead of a single spin; this may lead to exotic magnetic phases \cite{lee14b,lee14d,li15,wu15}. Also, while we have considered a classical drive here, it would be interesting to consider what happens with a quantum drive, i.e., the $\cos\omega t$ in Eq.~\eqref{eq:H_nonherm} would be replaced by creation and annihilation operators \cite{niemczyk10,forndiaz10}. The Hermitian quantum Rabi model was exactly solved recently \cite{braak11}, so perhaps the solution can be extended to the non-Hermitian case. 

\begin{acknowledgments}
We thank Peter Rabl for useful comments. This work was supported by NSF through a grant to ITAMP (T.L.) and through DMR-1054020 (Y.J.).
\end{acknowledgments}

\appendix

\section{Review of Salwen's perturbation theory} \label{sec:salwen}

Here, we review Salwen's perturbation theory (Sec.~3B of Ref.~\cite{salwen55}). Suppose we want to find the eigenvalues of a Hamiltonian $H$, but there are two nearly degenerate levels, $|a\rangle$ and $|b\rangle$. Salwen's method finds the effective Hamiltonian $H'$ for the subspace of $|a\rangle$ and $|b\rangle$. This effective Hamiltonian is a $2\times2$ matrix that captures the effect of all other levels on this subspace. $H'$ is written as a perturbation series of the off-diagonal elements.

Let $|\psi\rangle$ be an eigenvector of $H$ corresponding to eigenvalue $\epsilon$:
\begin{eqnarray}
H|\psi\rangle &=& \epsilon |\psi\rangle. \label{eq:H_app}
\end{eqnarray}
$|\psi\rangle$ can be expanded as
\begin{eqnarray}
|\psi\rangle &=& |a\rangle\langle a|\psi\rangle + |b\rangle\langle b|\psi\rangle + \sum_n |n\rangle\langle n|\psi\rangle, \label{eq:psi_app}
\end{eqnarray}
where the index $n$ runs over states other than $|a\rangle$ and $|b\rangle$. By plugging Eq.~\eqref{eq:psi_app} into Eq.~\eqref{eq:H_app}, we get
\begin{eqnarray}
\langle a|H|a\rangle\langle a|\psi\rangle + \langle a|H|b\rangle\langle b|\psi\rangle + \sum_{n}\langle a|H|n\rangle\langle n|\psi\rangle \quad\quad\quad\nonumber\\
=\epsilon\langle a|\psi\rangle, \quad\quad \label{eq:a_app}\\
\langle b|H|a\rangle\langle a|\psi\rangle + \langle b|H|b\rangle\langle b|\psi\rangle + \sum_{n}\langle b|H|n\rangle\langle n|\psi\rangle \quad\quad\quad\nonumber\\
=\epsilon\langle b|\psi\rangle, \quad\quad \label{eq:b_app}\\
\langle n|H|a\rangle\langle a|\psi\rangle + \langle n|H|b\rangle\langle b|\psi\rangle + \sum_{n'}\langle n|H|n'\rangle\langle n'|\psi\rangle \quad\quad\quad\nonumber\\
=\epsilon\langle n|\psi\rangle, \quad\quad \label{eq:n_app}
\end{eqnarray}
We can solve Eq.~\eqref{eq:n_app} for $\langle n|\psi\rangle$ in terms of $\langle a|\psi\rangle$ and $\langle b|\psi\rangle$:
\begin{eqnarray}
\langle n|\psi\rangle=\frac{A_n(\epsilon)\langle a|\psi\rangle + B_n(\epsilon)\langle b|\psi\rangle}{\epsilon-\langle n|H|n\rangle}, \label{eq:nsol_app}
\end{eqnarray}
where $A_n(\epsilon)$ and $B_n(\epsilon)$ depend on $\epsilon$ and the matrix elements of $H$. If $H$ is infinite-dimensional (as it is for a Floquet Hamiltonian), then we write $A_n(\epsilon)$ and $B_n(\epsilon)$ as power series in the off-diagonal elements of $H$, keeping enough terms to obtain the desired accuracy.

By plugging Eq.~\eqref{eq:nsol_app} into Eqs.~\eqref{eq:a_app} and \eqref{eq:b_app}, we obtain the effective $2\times2$ Hamiltonian for $|a\rangle$ and $|b\rangle$:
\begin{eqnarray}
H'&=&\left(\begin{array}{cc}
\langle a|H|a\rangle + V_{aa}(\epsilon)  &    V_{ab}(\epsilon)                      \\
V_{ba}(\epsilon)                         &  \langle b|H|b\rangle + V_{bb}(\epsilon)
\end{array}\right),
\end{eqnarray}
where
\begin{eqnarray}
V_{aa}(\epsilon)&=&\sum_n \frac{\langle a|H|n\rangle A_n(\epsilon)}{\epsilon - \langle n|H|n\rangle}\\
V_{bb}(\epsilon)&=&\sum_n \frac{\langle b|H|n\rangle B_n(\epsilon)}{\epsilon - \langle n|H|n\rangle}\\
V_{ab}(\epsilon)&=&\langle a|H|b\rangle + \sum_n \frac{\langle a|H|n\rangle B_n(\epsilon)}{\epsilon - \langle n|H|n\rangle}\\
                &=& V_{ba}^*(\epsilon)
\end{eqnarray}

In practice, the procedure is as follows. We write out Eq.~\eqref{eq:n_app}, then solve it for $\langle n|\psi\rangle$, thereby obtaining $A_n(\epsilon)$ and $B_n(\epsilon)$ as power series of the off-diagonal elements to desired order. Then we find $H'$ by calculating $V_{aa},V_{bb},V_{ab}$. We diagonalize $H'$ to find its eigenvalue $\epsilon$. However, if $H'$ itself depends on $\epsilon$, we  solve the resulting implicit equation for $\epsilon$.

\section{Boundaries for three-photon resonance} \label{sec:app_three}

Here, we calculate the boundaries for the three-photon resonance ($n=1$) beyond just a single line. This requires going to high enough order in perturbation theory. The effective Hamiltonian when $\omega_o\approx 3\omega$ is:
\begin{eqnarray}
H_F'&=&\left(\begin{array}{cc}
\frac{\omega_o}{2} + \delta_a   &   u                         \\
u                               &  -\frac{\omega_o}{2} + 3\omega + \delta_b
\end{array}\right),
\end{eqnarray}
where the level shifts of the two states $(\delta_a,\delta_b)$ can be different. The eigenvalues are
\begin{eqnarray}
\epsilon&=&\frac{3\omega+\delta_a+\delta_b}{2}\pm\frac{1}{2}\sqrt{4u^2+(\delta_a-\delta_b-3\omega+\omega_o)^2}. \nonumber\\ \label{eq:epsilon_3pho}
\end{eqnarray}

For convenience, we define a small parameter $\alpha$ to do perturbation theory in. We rewrite $\lambda,\epsilon,\omega$ in terms of $\alpha$:
\begin{eqnarray}
\lambda&=&\lambda' \alpha, \label{eq:app1}\\
\epsilon-\frac{\omega_o}{2}&=&\epsilon' \alpha^2,\\
\omega-\frac{\omega_o}{3}&=&\Delta \alpha^2,
\end{eqnarray}
where $\lambda',\epsilon',\Delta$ are on the order of unity. Using Salwen's perturbation theory, we find:
\begin{eqnarray}
\delta_a &=& -\frac{9\lambda^2}{4\omega_o}\alpha^2 - \frac{9\lambda'^2(9\lambda'^2+6\Delta\omega_o-10\epsilon'\omega_o)}{32\omega_o^3}\alpha^4,\quad\\
\delta_b &=& \frac{9\lambda^2}{4\omega_o}\alpha^2 + \frac{9\lambda'^2(9\lambda'^2-24\Delta\omega_o+10\epsilon'\omega_o)}{32\omega_o^3}\alpha^4,\quad\\
u&=&\frac{9i\lambda'^3}{4\omega_o^2}\alpha^3, \label{eq:app2}
\end{eqnarray}
where $\delta_a,\delta_b$ are accurate to $O(\alpha^4)$.

We plug Eqs.~\eqref{eq:app1}--\eqref{eq:app2} into Eq.~\eqref{eq:epsilon_3pho} and solve for $\epsilon'$. We want to find when $\epsilon'$ becomes complex, since that corresponds to the $\mathcal{PT}$ transition. This occurs when:
\begin{eqnarray}
0 &=& (3\lambda'^2+2\Delta\omega_o)^2 \nonumber\\
&& \quad + \frac{9\lambda'^2(5\lambda'^2+4\Delta\omega_o)(11\lambda'^2+2\Delta\omega_o)}{8\omega_o^2}\alpha^2, \label{eq:boundary_3pho}
\end{eqnarray}
which can be solved for $\lambda'$ in terms of $\Delta$. This agrees well with numerics for small $\lambda$ (see blue dotted lines in Fig.~\ref{fig:zoomin}). Note that if we omitted the second term in Eq.~\eqref{eq:boundary_3pho}, we would recover Eq.~\eqref{eq:line_multi}, since that is the lowest-order result.

\bibliography{rabi}

\end{document}